# Kondo holes in the 2D itinerant Ising ferromagnet Fe$_3$GeTe$_2$

Mengting Zhao[1,2], Bin-Bin Chen[3], Yilian Xi[1,2], Yanyan Zhao[4], Hongrun Zhang[1], Haifeng Feng[1,2*], Jincheng Zhuang[1], Xun Xu[1,2], Weichang Hao[1,3], Wei Li[1, 3,5*], Si Zhou[2,4], Shi Xue Dou[1,2], Yi Du[1,2*]

[1]BUAA-UOW Joint Research Centre and School of Physics, Beihang University, Beijing 100191, China

[2]Institute for Superconducting and Electronic Materials, Australian Institute for Innovative Materials, University of Wollongong, Wollongong, New South Wales 2500, Australia

[3]Key Laboratory of Micro-Nano Measurement-Manipulation and Physics (Ministry of Education), Beihang University, Beijing 100191, China

[4]Key Laboratory of Materials Modification by Laser, Ion and Electron Beams, Dalian University of Technology, Dalian 116024, China

[5]International Research Institute of Multidisciplinary Science, Beihang University, Beijing 100191, China

**Abstract**

Heavy fermion (HF) states emerge in correlated quantum materials due to the interplay between localized magnetic moments and itinerant electrons, but rarely appear in 3$d$-electron systems due to high itinerancy of $d$-electrons. Here, an anomalous enhancement of Kondo screening is observed at the Kondo hole of local Fe vacancies in Fe$_3$GeTe$_2$ which is a recently discovered 3$d$-HF system featuring of Kondo lattice and two-dimensional itinerant ferromagnetism. An itinerant Kondo-Ising model is established to reproduce the experimental results which provides insight of the competition between Ising ferromagnetism and Kondo screening. This work explains the microscopic origin of the $d$-electron HF states and inspires study of the enriched quantum many-body phenomena with Kondo holes in Ising ferromagnets.

The entanglement and hybridization of local magnetic moments and itinerant electrons lead to the emergence of exotic quantum phases and phenomena, including quantum criticality, topological order, and heavy fermion metallic behaviour in correlated quantum materials [1-6]. Among others, the HF states in a Kondo lattice model have attracted long-term interest, where the localized moment at each site is screened by the surrounding itinerant electrons and forms a spin singlet. Consequently, significant modulation of the Fermi surface volume and a dramatic increase in the electron effective mass occur when the screening effect appears in the Kondo lattice coherently [7,8]. Recently, a robust itinerant ferromagnetism with strong out-of-plane anisotropy has been confirmed in a two-dimensional (2D) van der Waals (vdW) $Fe_3GeTe_2$ (FGT) ferromagnet, even down to monolayers [9-14]. It shows Ising ferromagnetic characteristics with layer-number dependent magnetic phenomena. Remarkably, both HF and Kondo lattice behaviour were observed in this itinerant 2D Ising ferromagnet [15], which is scarce in the 3$d$-electron systems [16-18]. This suggests the coexistence of Ising ferromagnetism and electron itinerancy in FGT, despite a concrete microscopic modelling of the interplay between the two, is absent, and thus, the origin of the emergent HF states in the 2D Ising ferromagnet remains elusive.

Magnetic vacancies in the Kondo lattice system, known as Kondo holes, are of significance for engineering the HF states by tuning the hybridization between itinerant and localized electrons. Their impacts on the atomic heterogeneity of hybridization and microscopic characteristics of HF states can provide in-depth insight into the complex interactions in Kondo lattice systems [19-21]. In FGT, the Curie temperature, magnetic anisotropy, and net magnetization are decreased in the presence of Fe vacancies [22-25]. These vacancies lead to atomic-scale spatial variations in magnetic interactions. In addition, the electronic hybridization of localized and itinerant $d$-electrons is also expected to be modulated by Fe vacancies. Consequently, the Ising magnetic order and Kondo screening in FGT may vary in the vicinity of Fe vacancies, providing an intricate probe of the Kondo-lattice behaviour and HF states in this system.

Here, we reveal the role of intrinsic Fe vacancies as Kondo holes, and propose a Kondo-Ising model to understand the origin of HF states in FGT. Unusual modulation of hybridization strength due to anomalous enhancement of Kondo screening in the vicinity of each Fe vacancy site has been visualized by measuring the local density of states (LDOS). Through a comprehensive theoretical analysis using the density functional theory (DFT) and density matrix renormalization group (DMRG) methods, we establish the Kondo-Ising model for HF states in this Ising ferromagnets. We further reveal a concrete scenario for explaining the intriguing interplay between itinerant electrons, local moments, and Kondo holes in the 2D itinerant Ising ferromagnet. The Kondo holes weaken the local magnetic moment by breaking some nearest-neighbouring (NN) "Ising bonds" and concomitantly enhance the Kondo screening around it by reducing Pauli exclusion and relieving spin frustration between itinerant electrons. The agreement between experimental observations and the Kondo-Ising model results reveal the intricate role that the Fe vacancy plays in the Ising magnet and confirm FGT as a Kondo-lattice material with HF states.

The surface of FGT was first characterized by scanning tunneling microscopy (STM) and scanning tunneling spectroscopy (STS) at 4.2 K. A large-area, atomically flat surface is shown in **Fig. 1a**. **Fig. 1b** shows a typical terrace with the height of 8.2 Å, which corresponds to $c/2$ of the FGT lattice. In the $dI/dV$ spectrum (**Fig. 1c**), two peaks located at -180 mV and -50 mV can be assigned to the ferromagnetic (FM) states of Fe $3d$-states, while the shoulder peak at around 20 mV is attributed to the Kondo resonance (KR) state [15]. The STS features give rise to the coexistence of itinerant ferromagnetism and Kondo lattice behaviour in FGT. The evolution from the experimental acquired atomic structure by STM, through the simulated STM image to the crystal structure is shown in **Fig. 1e**, in which good agreement is reached. The brighter and less bright atoms can be assigned to Te and Fe(I) atoms (**Fig. 1e**), respectively, forming a hexagonal lattice with a lattice constant of 4.0 Å.

In the STM image, dark spots can be observed on the FGT surface, as circled in **Fig. 2a**. Although these dark spots are located at Te sites, the height variation between them and the surrounding Te atoms is only 24 pm, which is much smaller than the diameter of a single Te atom (Supplemental Material **Fig. S1**) [26]. Previous studies found that intrinsic Fe vacancies commonly form in FGT single crystal [22-25]. As Fe(II) sites are right beneath the topmost Te atoms, these dark spots may be reflecting Fe(II) vacancies. We compared $dI/dV$ spectra acquired at the dark spots and surrounding area, as shown in **Fig. 2b** and **2c**. The peaks corresponding to ferromagnetic (FM) states (FM1 and FM2) demonstrate an obvious reduction in intensity, which reflects the lesser contribution to the LDOS from Fe atoms. The unoccupied-state LDOS, in contrast, increases when the tip approaches the dark spot and indicates a local hole-like doping effect (Supplemental Material **Fig. S2**) [26,38,39]. As the Fe vacancies lead to the local deficiency of electrons, the dark spots in the STM images can be assigned to be Fe vacancies at Fe(II) sites. Moreover, the isolated Fe(II) vacancies should not carry magnetic moment, so that they can be regarded as Kondo holes in the FGT Kondo lattice. **Fig. 2c** is a line profile consisting of 30 spectra collected along the dashed line in **Fig. 2d**, which show the spatial distribution of the LDOS of FM as well as Kondo resonance (KR) states. Interestingly, the KR LDOS is enhanced at the Fe(II) vacancy centre and gradually decreases in the vicinity away from the vacancy. This is contrary to intuition in the Kondo lattice model, because a decreased FM LDOS due to Fe(II) vacancies should depress the local magnetic moments and reduce the Kondo screening effect and likewise the KR LDOS. The anomalous enhancement in KR LDOS, thus, cannot be explained by the Kondo lattice model for conventional $f$-electron heavy fermion systems.

In order to quantify the impact of the Fe(II) vacancy on the Kondo screening effect, the KR peaks acquired in the Fe(II) vicinity are fitted by using the following Fano resonance line-shape equation [40].

$$\frac{dI}{dV} = A_0 + A_K \frac{(q+\varepsilon)^2}{1+\varepsilon^2}, \varepsilon = \frac{eV - \varepsilon_0}{\Gamma} \quad (1)$$

in which, $A_0$ is a constant, $A_K$ is the KR amplitude, $q$ is the Fano symmetry parameter, $V$ is the sample bias voltage, $\varepsilon_0$ is the energy position, and $\Gamma$ is the resonance half width at the half maximum [41]. The FM peaks are fitted by a Lorentzian line-shape function [15], as shown in **Fig. 3a**. It is found that the value of $q$ rises from 0.6 at a pristine site to 1.2 at the Fe(II) vacancy site (**Fig. 3b**). Together with the enhanced intensity of the KR state (**Fig. 3c**), this implies that enhanced Kondo screening occurred at the Fe(II) vacancy sites. The amplitude of the two FM states decreased as the tip approached closer to the Fe(II) vacancy centre, which is consistent with the large area $dI/dV$ mapping (Supplemental Material **Fig. S**3) [26]. With positive sample bias, electrons tunnel into the unoccupied states of FGT or into the KR which lies close to Fermi level ($E_F$), as illustrated in **Fig. 3d**. The Fano symmetry parameter $q$ is determined by the ratio of tunnelling probability between the localized KR ($t_k$) and the direct conducting sea ($t_c$). A higher probability of tunnelling to the KR state results in higher value of $q$ [42]. Unlike the well-studied Kondo impurities of 3$d$ metal atom, with a Kondo screening effect that can be decreased by weakened the impurity-substrate interactions [41], the Kondo lattice system in FGT with translational symmetry share the same Fermi sea consisted of $s/p$ electrons from Te and Ge atoms, along with delocalized 3$d$ electrons and 4$s$ electrons of Fe atom. Thus, the enhanced $q$ indicates an enhanced Kondo screening effect, rather than a change in the tunneling probability between the tip and the Fermi sea of the sample.

From a comparison of the electronic structures of pristine (**Fig. 4a**) and defective FGT (**Fig. 4b**) obtained in DFT calculations, we can see that the Fe(II) vacancy can induce significant hole doping effects (Supplemental Material **Fig. S**4) [26], agreeing with our STS and previous angle-resolved photoelectron emission spectroscopy (ARPES) results [25].

Furthermore, as marked in the **Fig. 4c** and **4d**, there is a notable reduction of spin moment from 2.4 μ$_B$ to 2.1 μ$_B$ for the Fe(I) sites adjacent to the Fe(II) vacancy, which matches well with the weakened FM states in the STS result. Owing to a prominent FM coupling between nearest neighbour (NN) Fe(I) ions (~ 9 meV) on the same layer, direct FM spin exchange between local moments should be included in the theoretical analysis of the Kondo-lattice model for FGT.

To explain the enhancement of the KR in the vicinity of Fe(II) vacancy, we introduce a Kondo-lattice type many-body microscopic model to fully cover the correlations and hybridization between localized and itinerant electrons. Considering the two key features of robust ferromagnetism along the *c* axis, and strong hybridization between local moments and itinerant electrons accounting for its large thermal mass and Kondo lattice behaviour [15,22,23,27,43], we write down the Kondo-Ising chain Hamiltonian as follows:

$$H = \sum_i J_{Ising} S_i^z S_{i+1}^z + \sum_i J_{Kondo} S_i^a \cdot s_i^b + \sum_i \left( t_\parallel c_i^\dagger c_{i+1} + h.c. \right). \quad (2)$$

The summation ($\sum$) is over all spin sites other than the hole site located in the middle of the Kondo-Ising chain; $S_i^a$ and $s_i^b$ denote the local moments and spins of the conducting electrons at the $i^{th}$ site, respectively. $t_\parallel$ is the hopping amplitude of the conduction electrons between NN sites, $J_{Kondo}$ represents the Kondo coupling between the local moment and the conducting electron at the same local site *i*, and $J_{Ising}$ represents the Ising coupling between the NN local moments. The $c_i(c_i^\dagger)$ represent the fermion annihilation (creation) operator. The Kondo-Ising chain constitutes a strongly correlated many-body system, and we employed DMRG methods to simulate its ground state. In the pristine FGT Kondo lattice the local moments align in a FM order and the spin of itinerant electrons partially screen them, showing the Kondo lattice behaviour simultaneously (**Fig. 4e**). When an Fe vacancy as a Kondo hole is introduced into the system multiple intriguing consequences emerge owing to the correlations and hybridization of *d*-electrons in the system, as shown in **Fig. 4f**. It was found that the expectation value of local moments adjacent to the "hole" are decreased, while the Kondo screening by the spin of itinerant electrons is enhanced, which well reproduce the experimental observations.

Besides the effects on spin correlations, the Kondo hole has also a strong influence on the charge distribution near the Kondo-hole site, which can be clearly seen in **Fig. 4g**. By transforming the charge distribution from real to momentum (*k*) space, a period-two charge density wave (CDW) pattern is recognized by the pronounced peak in the charge structure factor that emerges at $\pi$ in the vacancy Kondo-Ising system(**Fig. 4h**). It is noteworthy that a 2 × 2 superstructure independent of sample bias and tunnelling distance has been experimentally observed in the *dI/dV* mapping (Supplemental Material **Fig. S5**) [26]. Although the one-dimensional (1D) chain model in DMRG calculations is insufficient to imitate the 2D Fermi surface of FGT, the coincidence of the emergence of Fermi surface modification indicates that they very likely share a similar origin. Overall, the DMRG calculations show that the Kondo hole can modify not only the magnetic properties of FGT, including the enhanced Kondo screening of local moments and decreased spontaneous FM ordering, but also alter the Fermi surfaces and charge correlations of itinerant electrons.

The relationship between magnetism and Kondo screening in a limited number of *d*-electron systems has been a long-term puzzle. For FGT, the latest ARPES results suggest that local magnetic moments play an indispensable role in the ferromagnetic ordering, which does not fit into the conventional Stoner scenario for itinerant electron ferromagnetism [44]. Also, in **Fig. 4a**, our DFT results for the pristine FGT structure show both rather flat and dispersive bands in the *d*-electron structure, indicating the existence of both localized and itinerant *d*-electrons in the system. Meanwhile, the local moments in FGT interact with the NN ones in a more direct way – Ising exchange interaction, as confirmed by the DFT calculations, which is very different from the Ruderman–Kittel–Kasuya–Yosida behaviour in *f*-electron HF systems. On the other hand, our experimental results have clearly revealed a competing interaction between the ferromagnetism and the Kondo effect in FGT, based on their opposite tendency in relation to Kondo holes, which is similar to what occurs in *f*-electron HF systems despite the very different magnetic interactions.

The reason can be found in the Kondo-Ising model of FGT proposed in this work, as illustrated by the schematic diagrams based on the DMRG calculations (**Fig. 4e** and **4f**). The spins of the itinerant electrons align antiparallel to local moments and induce a Kondo screening of the local moments at each site of the system. The spins of itinerant electrons tend to align antiparallel between NN sites which is energetically more favourable due to the Pauli exclusion. As the local moments have a strong FM ordering, competition exists between FM and Kondo screening that actually leads to spin frustration. This spin frustration is locally relieved by the presence of a Kondo hole. As shown in **Fig. 4f**, the spins of itinerant electrons in the close vicinity of the Kondo hole can preferentially align parallel to the local moments forming the Ising FM order, which, in turn, enhances the Kondo screening effect on the adjacent local moments. Similar competition between the 3*d*-electron FM and the Kondo effect has been observed by STS in cobalt dimers [45] and quantum point contacts of Fe/Co/Ni transition metal, indicating that the way of exchange coupling of the 3*d*-electron magnetic moments is critical for forming the Kondo screening [46]. More importantly, due to the Ising interaction in FGT, stronger local Kondo screening due to the presence of Kondo holes can enhance the hybridization between the adjacent local moments and itinerant electrons, rather than simply destroying the coherence of the Kondo lattice as in *f*-electron systems, as exemplified in Th doped $URu_2Si_2$ [20, 21].

In summary, we have performed STM and STS measurements on FGT, with an emphasis on the effects of Fe vacancy, i.e., Kondo holes. Decreased ferromagnetism and enhanced

Kondo screening in the vicinity of a Kondo hole were revealed and explained. By combining the experimental results with DFT and DRMG calculations, a concrete Kondo-Ising scenario for the strongly correlated *d*-electrons in FGT has been established, which can not only describe the competing relationship between the Kondo screening effect and Ising ferromagnetism, but also reveals the origin of the HF behaviour in FGT. This clarification of the role of Fe vacancies as Kondo holes in ferromagnetic and HF behaviour in FGT can enrich our understanding of the ferromagnetism and electronic properties of the 2D Ising ferromagnet FGT and paves the way to their spintronic engineering and application.

**Notes**: The authors declare no competing financial interest.

**Acknowledgements**

This work was supported by the Australian Research Council (DP170101467, FT180100585 and LP180100722) National Key R&D Program of China (2018YFE0202700), and National Natural Science Foundation of China (Nos. 11874003, 11974036, 11834014, 11904015, 51672018) and Beijing Natural Science Foundation (Z180007). The work was partially supported by the UOW-BUAA Joint Research Centre and a Vice Chancellor's Research Fellowship from the University of Wollongong. The authors acknowledge the computer resources provided by the NCI National Facility through the University of Wollongong Partner Share Scheme, and the Supercomputing Centre of Dalian University of Technology. We thank Professor Lan Wang, Professor Jiabao Yi, Professor Lan Chen, and Professor Faxian Xiu for valuable discussions and communications, and Dr. Tania Silver for critical reading of the manuscript.

**References**

[1] N. D. Mathur, F. M. Grosche, S. R. Julian, I. R. Walker, D. M. Freye, R. K. W. Haselwimmer, and G. G. Lonzarich, Nature **394,** 39 (1998).

[2] Q. Si, S. Rabello, K. Ingersent, and J. L. Smith, Nature **413,** 804 (2001).

[3] P. Coleman, AJ. Schofield, Nature **433**, 226 (2005).

[4] G. R. Stewart, Rev. Mod. Phys. **73**, 797 (2001).

[5] A. V. Chubukov, C. Pépin, J. Rech, Phys. Rev. Lett. **92**, 147003 (2004).

[6] B. Shen, Y. Zhang, Y. Komijani, M. Nicklas, R. Borth, A. Wang, Y. Chen, Z. Nie, R. Li, X.Lu, H. Lee, M. Smidman, F. Steglich, P. Coleman and H. Yuan, Nature **579,** 51 (2020).

[7] E. Pavarini, E. Koch, and P. Coleman, Forschungszentrum Jülich, **5**, 21 (2015).

[8] Q. Si, F. Steglich, Science **329** 1161 (2010).

[9] Z. Fei, B. Huang, P. Malinowski, W. Wang, T. Song, J. Sanchez, W. Yao, D. Xiao, X. Zhu, A. F. May, W. Wu, D. H. Cobden, J-H. Chu, and X. Xu, Nat. Mater. **17,** 778 (2018).

[10] Y. Deng, Y. Yu, Y. Song, J. Zhang, N. Z. Wang, Z. Sun, Y. Yi, Y. Z. Wu, S. Wu, J. Zhu, J. Wang, X. H. Chen, and Y. Zhang, Nature **563**, 94 (2018).

[11] X. Wang, J. Tang, X. Xia, C. He, J. Zhang, Y. Liu, C. Wan, C. Fang, C. Guo, W. Yang, Y. Guang, X. Zhang, X. Zhang, H. Xu, J. Wei, M. Liao, X. Lu, J. Feng, X. Li, Y. Peng, H. Wei, R. Yang, D. Shi, X. Zhang, Z. Han, Z. Zhang, G. Zhang, G. Yu, and X. Han, Sci. Adv. **5**, eaaw8904 (2019).

[12] S. Liu, X. Yuan, Y. Zou, Y. Sheng, C. Huang, E. Zhang, J. Ling, Y. Liu, W. Wang, C. Zhang, J. Zou, K. Wang, and F. Xiu, npj 2D Mater. Appl. **1,** 30 (2017).

[13] C. Tan, J. Lee, S.-G. Jung, T. Park, S. Albarakati, J. Partidge, M. R. Field, D. G. McCulloch, L. Wang, and C. Lee. Nat Commun. **9,** 1554 (2018).

[14] M. Alghamdi, M. Lohmann, J. Li, P. R. Jothi, Q. Shao, M. Aldosary, T. Su, B. P. T. Fokwa and J. Shi. Nano Lett. **19**, 4400 (2019).

[15] Y. Zhang, H. Lu, X. Zhu, S. Tan, W. Feng, Q. Liu, W. Zhang, Q. Chen, Y. Liu, X. Luo, D. Xie, L. Luo, Z. Zhang, and X. Lai, Sci. Adv. **4**, eaao6791 (2018).

[16] C. Urano, M. Nohara, S. Kondo, F. Sakai, H. Takagi, T. Shiraki, and T. Okubo Phys. Rev. Lett. **85**, 1052 (2000).

[17] Y. P. Wu, D. Zhao, A. F. Wang, N. Z. Wang, Z. J. Xiang, X. G. Luo, T. Wu, and X. H. Chen. Phys. Rev. Lett. **116**, 147001(2016).

[18] H. Kotegawa, M. Matsuda, F. Ye, Y. Tani, K. Uda, Y. Kuwata, H. Tou, E. Matsuoka, H. Sugawara, T. Sakurai, H. Ohta, H. Harima, K. Takeda, J. Hayashi, S. Araki, and T. C. Kobayashi. Phys. Rev. Lett. **124**, 087202(2020).

[19] J. D. Thompson, Proc. Natl. Acad. Sci. **108**, 18191 (2011).

[20] M. H. Hamidian, A. R. Schmidt, I. A. Firmo, M. P. Allan, P. Bradley, J. D. Garrett, T. J. Williams, G. M. Luke, Y. Dubi, A. V. Balatsky, and J. C. Davis, Proc. Natl. Acad. Sci. **108**, 18233 (2011).

[21] A. R. Schmidt, M. H. Hamidian, P. Wahl, F. Meier, A. V. Balatsky, J. D. Garrett, T. J. Williams, G. M. Luke and J. C. Davis, Nature **465**, 570 (2010).

[22] V. Y. Verchenko, A. A. Tsirlin, A. V. Sobolev, I.A. Presniakov, and A. V. Shevelkov, Inorg. Chem. **54**, 17, 8598 (2015).

[23] A. F. May, S. Calder, C. Cantoni, H. Cao, and M. A. McGuire, Phys. Rev. B **93**, 014411 (2016).

[24] K. Kim, J. Seo, E. Lee, K.-T. Ko, B. S. Kim, B. G. Jang, J. M. Ok, J. Lee, Y. J. Jo, W. Kang, J. H. Shim, C. Kim, H. W. Yeom, B. Il Min, B-J. Yang and Jun Sung Kim Nat. Mater. **17,** 794 (2018).

[25] S. Y. Park, D. S. Kim, Y. Liu, J. Hwang, Y. Kim, W. Kim, J.-Y. Kim, C. Petrovic, C. Hwang, S.-K. Mo, H. Kim, B.-C. Min, H. C. Koo, J. Chang, C. Jang, J. W. Choi, and H. Ryu, Nano Lett. **20**, 95 (2020).

[26] Supporting Information, contains the sample preparation, details of the experimental methods, DFT and DMRG calculations. Including Ref 27-37.

[27] B. Chen, J. Yang, H. Wang, M. Imai, H. Ohta, C. Michioka, K. Yoshimura, M. Fang, J. Phys. Soc. Jpn. **82**, 124711 (2013).

[28] I. Horcas, R. Fernandez, J.M. Gomez-Rodriguez, J. Colchero, J. Gomez-Herrero and A. M. Baro, Rev. Sci. Instrum. **78**, 013705 (2007).

[29] G. Kresse and J. Furthmüller, Phys. Rev. B **54**, 11169 (1996).

[30] J. P. Perdew, K. Burke, and M. Ernzerhof, Phys. Rev. Lett. **77**, 3865 (1996).

[31] G. Kresse and D. Joubert Phys. Rev. B **59**, 1758 (1999).

[32] G. Henkelman, A. Arnaldsson, and H. J´onsson, Comput. Mater. Sci. **36**, 354 (2006).

[33] S. R. White, 1992 Phys. Rev. Lett. **69**, 2863 (1992).

[34] S. R. White, Phys. Rev. B **48**, 10345 (1993).

[35] U. Schollwöck, Rev. Mod. Phys. **77**, 259 (2005).

[36] C. C. Yu and S. R. White, Phys. Rev. Lett. **71**, 3866 (1993).

[37] A. Weichselbaum, Ann. Phys. **327** 2972 (2012).

[38] D. Wong, J. V. Jr, L. Ju, J. Lee, S. Kahn, H. Tsai, C. Germany, T. Taniguchi, K. Watanabe, A. Zettl, F. Wang, and M. F. Crommie, Nat. Nanotech. **10**, 949(2015).

[39] Y. Wang, V. W. Brar, A. V. Shytov, Q. Wu, W. Regan, H. Tsai, A. Zettl, L. S. Levitov, and M. F. Crommie, Nat. Phys. **8**, 653 (2012).

[40] M. Moro-Lagares, R. Korytár, M. Piantek, R. Robles, N. Lorente, J. I. Pascual, M. R. Ibarra, and D. Serrate, Nat. Commun. **10**, 2211 (2019).

[41] A. F. Otte, M. Ternes, K. von Bergmann, S. Loth, H. Brune, C. P. Lutz, C. F. Hirjibehedin, and A. J. Heinrich, Nat. Phys. **4,** 847 (2008).

[42] M. Ternes, A. J Heinrich, W-D. Schneider, J. Phys.: Condens. Matter **21**, 053001 (2009).

[43] J.-X. Zhu, M. Janoschek, D. S. Chaves, J. C. Cezar, T. Durakiewicz, F. Ronning, Y. Sassa, M. Mansson, B. L. Scott, N. Wakeham, E. D. Bauer, and J. D. Thompson. Phys. Rev. B **93**, 144404 (2016).

[44] X. Xu, Y. W. Li, S. R. Duan, S. L. Zhang, Y. J. Chen, L. Kang, A. J. Liang, C. Chen, W. Xia, Y. Xu, P. Malinowski, X. D. Xu, J.-H. Chu, G. Li, Y. F. Guo, Z. K. Liu, L. X. Yang, and Y. L. Chen, Phys. Rev. B **101**, 201104 (2020).

[45] W. Chen, T. Jamneala, V. Madhavan, and M. F. Crommie, Phys. Rev. B, **60**, R8529 (1999).

[46] M. R. Calvo, J. Fernández-Rossier, J. J. Palacios, D. Jacob, D. Natelson, and Carlos Untiedt, Nature **458,** 1150 (2009).

**Figures and Figures Captions**

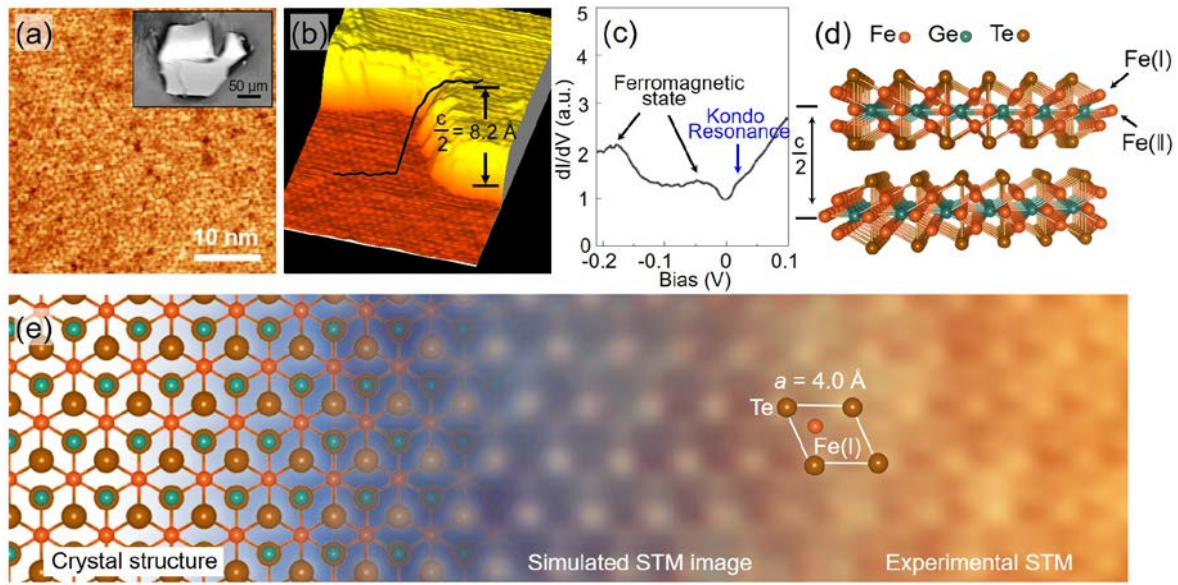

**Fig. 1.** Crystal and electronic structure of FGT. (a) Large-scale STM image of the surface of FGT. Insert: Optical image of FGT. (b) Atom-resolved STM topography ( sample bias, 9 mV). (c) *dI/dV* curves for FGT . (d) Side views of the crystal structure of FGT. (e) Schematic diagram of the evolution from the atomic structure of FGT to the simulated STM image and then to the experimental STM image (40 mV) (from left to right).

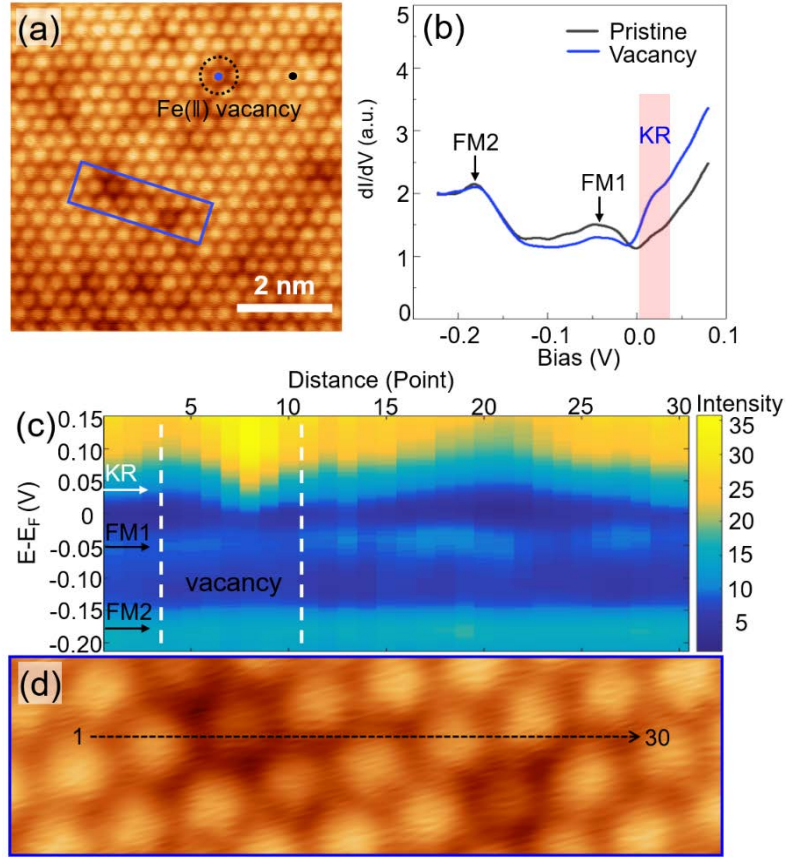

**Fig. 2.** Fe(II) vacancy characteristics in the FGT. (a) Atomic-resolution STM image of FGT with Fe(II) vacancies. The black dotted circle identifies the Fe(II) vacancies. (b) A pair of *d*I/*d*V spectra taken over areas with and far away from vacancy. (c) Waterfall image of 30 successive *d*I/*d*V curves measured along the dashed line in (d). (d) Zoomed STM image of the area marked by the blue rectangle in a.

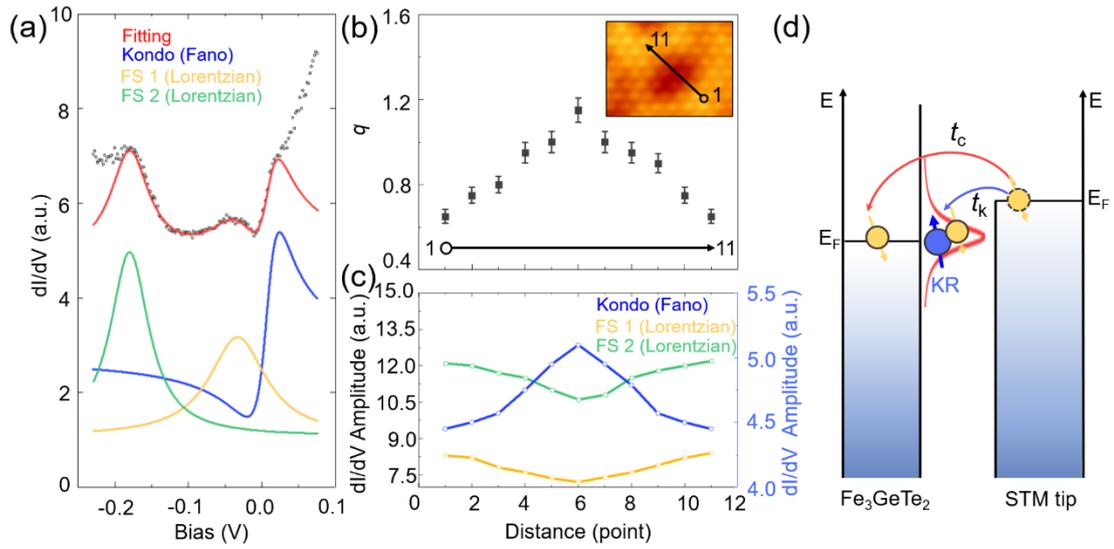

**Fig. 3.** Analysis of the STS curves of FGT. (a) Fitting of experimental *dI/dV* curves. The blue curve is a Fano fit with the best fit parameters of $q = 1.40 \pm 0.04$, $\varGamma = 26 \pm 5$ mV. (b) and (c) the obtained $q$ and amplitude depend on the distance from the vacancy. (d) Schematic illustration of the tunnelling process in the STS with electrons marked as yellow circle with arrow.. The blue circle with an arrow represents the local moment of an Fe atom.

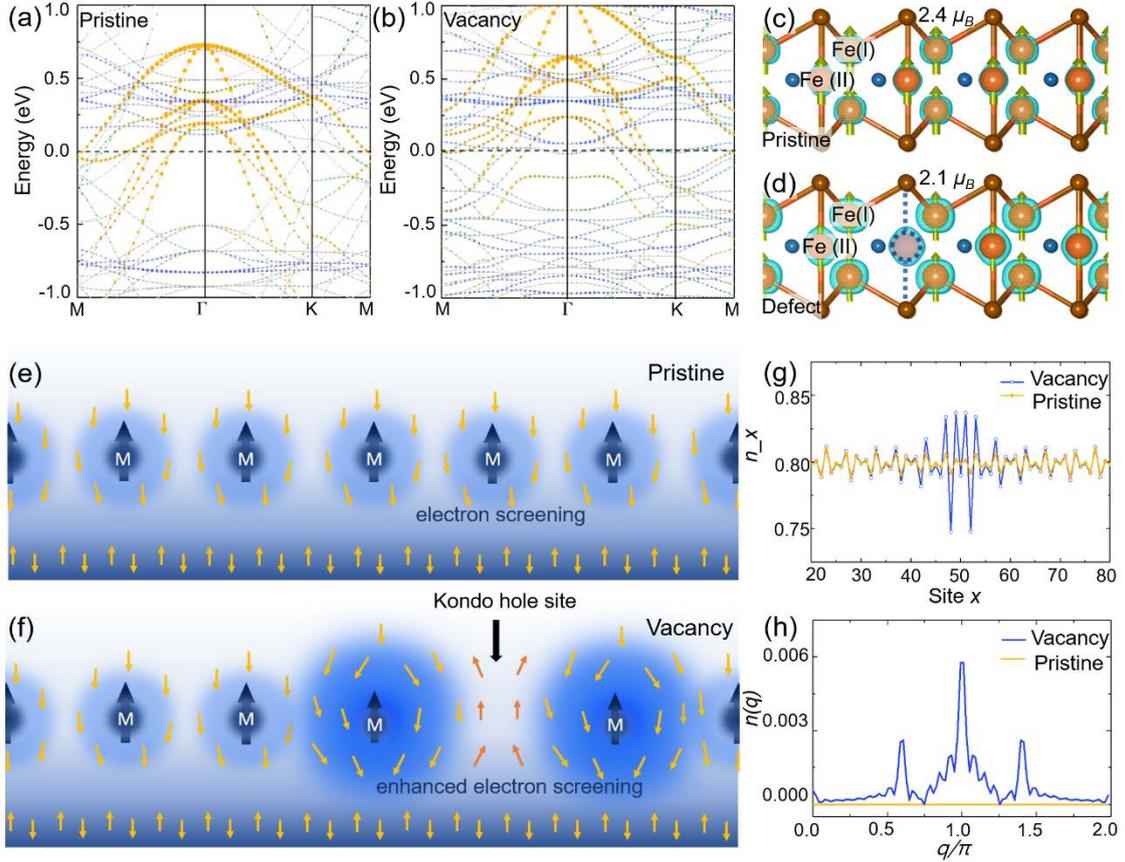

**Fig. 4.** Theoretical simulation of the effects of Fe(II) vacancies on the FGT. (a) and (b) are electronic structures for pristine FGT and FGT with vacancies, respectively. (c) and (d) are the side views of the crystal structures for these two situations. The blue atmosphere around the Fe atom represents the charge distribution. (e) and (f) are schematic diagrams of the Kondo screening in a pristine Kondo lattice and in a lattice with a Kondo hole, respectively. (g) and (h) are the real space charge density distribution and the corresponding momentum-space structures, respectively. n($q$) is the Fourier transform of the real-space distribution n$_x$ at site $x$.